\documentclass{article}
\usepackage{graphicx}
\usepackage{tablefootnote}
\usepackage{indentfirst,csquotes}

\usepackage{amssymb,amsthm,amsmath}
\usepackage{xcolor,paralist,hyperref,titlesec,fancyhdr,etoolbox}

\usepackage[inkscapelatex=false]{svg}

\hypersetup{ colorlinks=true, linkcolor=black, filecolor=black, urlcolor=black }

\usepackage{lipsum}

\title{End-to-End Demonstration for CubeSatellite Quantum Key Distribution} %%%%%%%%%%%%
\author{Peide Zhang*$^1$, Jaya Sagar$^2$, Elliott Hastings$^2$, Milan Stefko$^1$, Siddarth Joshi$^1$, John Rarity$^1$}

\begin{document}

\maketitle

\begin{abstract}
Quantum key distribution (QKD) provides a method of ensuring security using the laws of physics, avoiding the risks inherent in cryptosystems protected by computational complexity. Here we investigate the feasibility of satellite-based quantum key exchange using low-cost compact nano-satellites. This paper demonstrates the first prototype of system level quantum key distribution aimed at the Cube satellite scenario. It consists of a transmitter payload, a ground receiver and simulated free space channel to verify the timing and synchronisation (T\&S) scheme designed for QKD and the required high loss tolerance of both QKD and T\&S channels. The transmitter is designed to be deployed on various up-coming nano-satellite missions in the UK and internationally. The effects of channel loss, background noise, gate width and mean photon number on the secure key rate (SKR) and quantum bit error rate (QBER) are discussed. We also analyse the source of QBER and establish the relationship between effective signal noise ratio (ESNR) and noise level, signal strength, gating window and other parameters as a reference for SKR optimization. The experiment shows that it can tolerate the 40 dB loss expected in space to ground QKD and with small adjustment decoy states can be achieved. The discussion offers valuable insight not only for the design and optimization of miniature low-cost satellite-based QKD systems but also any other short or long range free space QKD on the ground or in the air.
\end{abstract} 

\bigskip
{\let\thefootnote\relax

\footnotetext{[1] Department of Engineering, University of Bristol, BS8 1UB, Bristol, UK}
\footnotetext{[2] Department of Physics, University of Bristol, BS8 1UB, Bristol, UK}
\footnotetext{email: peide.zhang@bristol.ac.uk}

\footnotetext{\textbf{Abbreviations:} QKD, quantum key distribution; SKR, security key rate; QBER, quantum bit error rate; ESNR, effective signal noise ratio; OGS, optical ground station; HDBC, hybrid de Bruijn code; LEO, low earth orbit; GEO, geosynchronous earth orbit; GPS, global positioning system; PPS, pulse per second; GBP, gain–bandwidth product; SNR, signal noise ratio; PRBS, pseudorandom binary sequence; SWaP, size, weight and power; EMC, electromagnetic compatibility; WCP, weak coherent pulse; PNS, photon number splitter;  SPCM, single-photon counting module; PBS, polarising beam splitter; FOV, field of view; APD, avalanche photodiodes; TDC, time to digital convert; FPGA, field programmable gate arrays; FWHM, full width half maximum; NEP, noise equivalent power; DCR, dark count rate; SNSPD, superconducting nanowire single-photon detector; NSKR, normalised SKR; SKL, security key length.}

}

\section{Introduction}

Quantum technology has already started a technological revolution in the fields of communication~\cite{2020_Pirandola}, sensing~\cite{2018_Proctor, 2020_Guo} and computing due to its enhanced capabilities. Quantum networks are moving step by step from the laboratory to end-user applications due to the threat of emerging quantum computers to existing cryptographic systems and the quest for unconditional security. A main difficulty to establish a global scale quantum network is exponential loss in optical fibres which limits the direct communication distance to around 1000 km~\cite{2016_Yin, 2020_Chen, 2020_Boaron, 2023_Liu, 2018_Wang}. Quantum repeaters are a good option for extending communication distances, but global extension through a large number of repeaters is not realistic~\cite{2021_Mustafa}. In addition, establishing fiber-based networks in areas isolated from the fibre backbone is impractical due to geographical factors and the absence of infrastructure, while mobile platforms require more flexibility of communication locations.

The near vacuum and microgravity environment of space makes it a potential way to test fundamental physics, including quantum technology~\cite{2022_Belenchia}. Early feasibility studies showed that satellite-based quantum communications can transmit photons over distances of several thousand kilometres \cite{2002_Rarity, 2003_Aspelmeyer}, making it ideal to create  worldwide key distribution by linking into fibre QKD networks for local distribution~\cite{2021_Chen}. The idea was discussed for several years and the interest reached a peak with the successful launch and demonstration of the Micius satellite~\cite{2017_Liao}. This first demonstration was a large and complex satellite weighing some 500kg requiring a dedicated, thus expensive, launch. As technology advances, there is an expectation that in-orbit demonstrations of quantum key distribution using nanosatellites will be possible due to their compact design, lower cost and the feasibility of establishing constellations~\cite{2020_Villar}. 

However, the limited space and scarce power budget raise a series of challenges for the payload implementation. A Cube-Satellite's size is from 1 U to 16 U\footnote{1U is a standard dimension (Units or “U”) of 10 cm x 10 cm x 10 cm}, which requires the payload to be compact. It is a particular challenge for the optical component miniaturisation, electronic integration and electromagnetic compatibility design. The solar panel size and the corresponding generated electric power decreases with a smaller satellite, which requires a more elegant power distribution and higher efficiency device selection. This limited resource exacerbates the difficulty of acquisition, pointing and tracking (APT), as well as the T\&S. 

In this work, we built up an end-to-end testbench used for demonstrating satellite-based QKD performance in the laboratory. The system used weak coherent pulse as quantum source. The T\&S between the satellite and optical ground station (OGS) is achieved using a hybrid de Bruijn code (HDBC) implemented by a high power pulsed laser diode ~\cite{2021_Zhang}. In the free space test channel, we use a variable ND filter to simulate the channel loss changing when the satellite is in orbit. We perform a series experiments to simulate the scenario in which the channel loss, and background noise is changing. Also we investigate the effect on secure key generation when selecting different gate widths in the postprocessing and the finite key length that can be generated. The work described here is the proof of principle experiment which reproduces the losses, and background noise expected in a space channel as modelled in a separate paper \cite{2020_Mazzarella} using a robust synchronisation scheme ~\cite{2021_Zhang} that enables operation in high loss and automatically corrects the varying Doppler shifts between the clocks. We are able to optimise gate widths and photon numbers to exchange keys over high loss links using electronics and hardware which we are now developing to fit a $\frac{1}{2}$U system \cite{2023_Jaya}. The transmitter and post processing software developed here is planned to be implemented in the UK Quantum communications hub SPOQC Mission~\cite{SPOQCWeb} where collaborators are developing a 12U nano-satellite with telescope~\cite{2023_Simmons} and full acquisition, pointing and tracking. The success of this work has also led to further planned missions including the UK-Canada add-on QEYSSAT mission~\cite{CanadaWeb} and the ROKS mission ~\cite{CraftPWeb}.

\section{Challenge and System Design}
\label{sec_2}
 
Although space-based QKD is a very promising solution for long-range quantum communication on a global scale, the system design faces various security and technical challenges, which are even more pronounced when using compact nanosatellite platforms with constrained resources. The miniaturization of satellites is a trend, and the accompanying power consumption and space reduction put higher requirements on the conversion efficiency of the devices and the design of the optical path. For example, a compact, simple-driven, and efficient single-mode beacon light source is a strong requirement, which can satisfy synchronization, tracking, and polarization reference with sufficient margin. On the ground receiving end, in order to avoid the operational limitations of fixed sites and excessive investment, mobile small-aperture optical telescopes are highly regarded. But the problem it brings is that the receiving efficiency drops so that a higher loss budget needs to be met. The following two subsections describe the challenges faced by onboard QKD systems and the corresponding solutions.

\textbf{Link Efficiency}. Compared with the maximum communication distance around 1000 km of optical fiber~\cite{2023_Liu}, satellite QKD can realize communication ranges of several thousand kilometers. However, due to the beam diffraction limit and limited receiver aperture, both beacon and quantum beam suffer diffraction loss. In the SPOQC Mission for example, geometric losses can rise to 27.7 dB at the low elevation angles (10 degrees) and 500 km altitude when quantum beam divergence is specified as 10 microrad ($1/e^2$) ~\cite{2020_Mazzarella}. Here the downlink T\&S beacon is designed to be slightly larger divergence at 18.7 microrad ($1/e^2$) to reliably perform the timing and synchronisation, resulting in 5.4 dB increase in geometric losses compared to the quantum signal (ie up to 33.1 dB loss). For system efficiency, as the transmitter can be treated as a trusted node, it is only necessary to ensure that the mean photon number $\mu$ at the telescope's exit aperture meets the security conditions, which alleviates internal optical losses. However for the beacon, the transmitter internal loss is about 3 dB due to the alignment error and the transmission of the dichroic mirror for combining the beacon and quantum beam. Carbon dioxide, oxygen, nitrogen, water and aerosol/particulate species in Earth atmosphere determine a large variety of absorption and scattering processes. The transmission at 785 nm is 0.56 (2.5 dB loss) and at 905 nm is 0.96 (0.2 dB loss) in zenith.~\cite{2012_Sabatini, 2022_Giggenbach} and both are about 7.9 dB loss at a distance of 1700 km, corresponding to a 10 degrees elevation angle. The downlink beam configuration suffers comparatively little loss due to turbulence compared with the uplink beam configuration. The geometric losses depend on the OGS receiver telescope aperture area, and we assume in the Hub mission estimates that a telescope diameter of 70 cm is used. In the downlink scheme, only the end of the optical path (close to the surface of the Earth) is affected by the turbulence, so the pointing error resulting from turbulence will lead to some additional loss. Here we assume pointing error losses (including from turbulence) will contribute 3 dB loss and OGS internal filtering and coupling will contribute another 3.8 dB losses, and the detector efficiency of 60\% adds another 2.2 dB of loss.  So, the total quantum link loss is between 28.6 dB and 44.6 dB while the beacon link loss is between 31.7 dB and 50 dB. The loss budget is summarised in Table.~\ref{tab1} and the loss modelling is presented in the appendix.  

\begin{table}[ht]\scriptsize
\centering

\caption{\label{tab1}Loss budget comparison between two missions \tablefootnote{The minimum and maximum values of the loss correspond to the satellite position at 90 deg elevation angle and 10 deg elevation angle with 500 km orbit, respectively.}}

\begin{tabular}{@{}lllll}
\hline
 & ROKs Q (785 nm) & SPOQC Q(785 nm) & ROKs B (905 nm) & SPOQC B (915 nm)\\
\hline

Tx Telescope Diameter (exp) & 90 mm & 80 mm & 90mm & 80 mm\\

OGS Telescope Diameter (exp) & 432 mm & 700 mm & 432 mm & 700 mm\\

Divergence (sim) & 10 urad & 10 urad & 100 urad & 18.7 urad\\

Losses internal to transmitter (exp) & 0 dB & 0 dB & 13.2 dB\tablefootnote{In order to get enough peak power for reliable SNR in the receiver APD, the beacon source uses a surface mount multi-mode laser diode which could provide up to 100W peak power but could only get 0.28 mrad with a 5mm beam size. This will result in a larger divergence compared with the target 100urad. On the other hand, the limited internal aperture (3.6mm) before the transmitted telescope clips part of the beam, which introduce an extra loss.} & 3 dB\\

Geometric Losses in optical channel (sim) & 21.7 dB - 32.3 dB & 17.1 dB - 27.7 dB & 41.7 dB - 52.3 dB & 22.5 dB - 33.1 dB\\

Turbulence and pointing losses~\cite{2017_Liao} & 3 dB & 3 dB & 3 dB & 3 dB\\

Atmosphere Absorption (ref) & 2.5 dB - 7.9 dB & 2.5 dB - 7.9 dB & 0.17dB - 7.9dB & 0.17 dB - 7.9 dB\\

Losses in OGS (exp) & 3.8 dB & 3.8 dB & 3 dB & 3 dB\\

Quantum Detection efficiency (exp) & 2.2 dB & 2.2 dB & - & -\\
\hline
\textbf{Total Loss} & 33.2 dB - 49.2 dB & 28.6 dB - 44.6 dB & 61.1 dB - 79.4 dB& 31.7 dB - 50.0 dB\\
\hline

\end{tabular}

\end{table}

\textbf{Relative Motion}. Currently in-orbit and under-development QKD satellites are all running in low earth orbit (LEO). Unlike geosynchronous earth orbit (GEO) satellites, LEO orbit periods are much faster than the Earth's rotation period. For a LEO at 500 km altitude, its orbital period is about 1.2 hours. This results in continuous  relative motion between the satellite and the OGS, which leads, on the one hand to a constant variation in link loss and on the other hand to high demands on the tracking and pointing systems between the satellites and OGS. In addition, due to the varying attitude changes of the satellites relative to the OGS, polarisation compensation is required to correct the polarisation reference between both sides, which is particularly important in the case of the BB84 protocol based on polarisation coding. Last but not least, varying linear relative motion causes the signal to carry the Doppler shift, which not only increases signal jitter but also rises a wavelength shift for both beams. Assuming $\pm$10 km/sec maximum relative velocity this would imply $\pm3\times 10^{-5}$ relative frequency shift~\cite{1998_Ali}. A very narrow bandpass filter might cause transmission variability due to the wavelength shifting from the Doppler Effect, but this is negligible for a filter whose window is $>1$ nm. The effect of Doppler shift on temporal synchronisation using clock recovered from a 100 kHz beacon is also negligible compared with the 100 ps synchronisation jitter requirement. 

\textbf{Timing and Sync}. To ensure consistent timing between the transmitter and receiver, an elegant time synchronisation mechanism needs to be designed as large and costly devices such as high precision atomic clocks are not suitable for this scenario. Several different approaches have been utilized for T\&S over long free space channels, including using pulsed lasers with various modulation patterns, and even combined with global positioning system (GPS) pulse per second (PPS) as a hybrid scheme. However, the > 50 dB loss budget greatly increases the difficulty of precision beacon timing detection. A ultra-weak optical signal detection requires a detector with very high sensitivity and low noise level. A higher detecting bandwidth also provides a better timing jitter due to the sharper rise time. However, due to the limitations of the detector gain–bandwidth product (GBP), sensitivity and bandwidth need to be traded off to achieve the best SNR.  Additionally, a large variation of incident power can result in pulse detection failure due to signal saturation. More importantly, the amplitude variation of the converted electronic pulse causes large discrimination errors of up to nanosecond level when using constant threshold signal discrimination, which significant increase the timing jitter and degrades the SNR~\cite{2016_MeyerScott}. Our solution is detailed below and in previous papers~\cite{2021_Zhang, 2023_Zhang}.

\textbf{Resource Limits}. In the global space launch market, the cost of satellites is gradually decreasing but is still one of the main budget items. A great trend is using the 'carpool' launch mode to reduce launch costs significantly in which a group of nano-satellites share a rocket to be launched into space. The miniaturisation thus becomes an essential step in research and commercialisation. This requires careful optimisation to ensure operational requirements are met within a tight size, weight and power (SWaP) budget. Small size batteries and solar panels can only provide limited power which restricts the payload consumption while limited space and weight budgets place greater demands on system integration, which involves the use of lighter materials, better heat dissipation, and more reliable electromagnetic compatibility (EMC) design. Here we also need a compact optical design strong and stable enough to withstanding launch forces without misalignment.

\subsection{Communication Link Design}

Designing a satellite-based QKD system is not as easy as a fibre-based system due to the need to establish and maintain a stable free space channel and the high loss which requires certain  protocols to be implemented to ensure the security of QKD communications. The potential for eavesdroppers to intercept valid information forces us to limit the photon number in a single pulse less than 1 characterized by the mean photon number $\mu$. In practice, the Poisson distribution of a weak coherent pulse (WCP) sources requires setting the $\mu<0.1$ to ensure that the probability of multi-photon pulses is small, thus limiting the information gain of a potential eavesdroppers via multi-photon signal splitting to within the acceptable range~\cite{2015_Bourgoin}. In fact in high loss scenarios the maximum $\mu$ drops proportional to the loss significantly reducing the efficiency of key generation and limits the communication distance. To get around this photon number splitting (PNS) attack, decoy-state BB84 protocols have emerged~\cite{2008_Ma, 2003_Hwang}, allowing for higher communication efficiency and longer distances allowing secure key generation in high loss scenarios. According to the latest research, the $\mu$ can be as high as 0.8 while retaining full security ~\cite{2017_Liao}. From an engineering viewpoint, the implementation of decoy-states requires pulse energy control on each pulse, which can be implemented  using a variable current driver or dual-drivers to trigger the weak coherent pulse (WCP) source. In addition, any differences between pulses from the 4 laser diodes in temporal, spatial and wavelength domains potentially leaks information to Eve without polarisation measurement, which requires the emitted pulses to have the same pulse shape and emission time. This places high demands on signal path matching, bandwidth control and wavelength consistency in the design of the light source. A resonant capacitive discharge driver is normally used for driving the short pulse laser, leading to resonance effects in the circuit, with a series of fading post-pulses following the main pulse at the resonance frequency when under-damped. This is a potential loophole for Eve intercepting valid information without introducing extra noise.

The BB84 protocol encodes photons using four independent polarisation states, but the laser beam generated by an actual source can never reach 100\% linear polarization.  When photons coded as H-polarisation are incident on an ideal polarising beam splitter (PBS), a small fraction will inevitably  enter into the V polarisation channel and  thus contribute to a raw key error rate, denoted as $e_{SP}$. Similarly, the imperfection of practical PBS's introduce unavoidable QBER as well denoted as $e_{PBS}$. In practice, $e_{PBS}$ can be estimated based on the extinction ratio of the PBS. The $e_{SP}$ is decided by the laser and polarizer. A polarizer with higher extinction ratio gives a lower QBER. These two parts are only determined by the device performance and working environment such as temperature and humidity, rather than the system design parameters, so they are defined as intrinsic QBER, $e_{I}$.

\begin{equation}
e_{I} = F(e_{SP}, e_{PBS})
\label{eq:1}
\end{equation}

$F$ is the QBER combination function, which is defined in equation.~\ref{eq:2}. Assuming there are $n$ bits of binary variables (representing the polarisation state of photons) with random value. Part of the bits are flipped randomly in the first operation and another part of the bits are flipped randomly in a second operation. As long as the two operations sample randomly and independently, the actual flipped bits after two rounds can be calculated according to the inclusion-exclusion principle.

\begin{equation}
F(e_1, e_2) = e_1 + e_2 - 2e_1e_2
\label{eq:2}
\end{equation}

When propagating in the atmosphere, the scattering and refraction of the suspended particles results in the depolarization, which further increases the QBER. Another significant point is the background light noise within the telescope receiving field of view (FOV), especially in systems based on large-diameter telescopes. Since sun radiation contains a large portion energy in the near-infrared band, it deteriorates or even fails the QKD, which is also the reason why most of QKD demonstrating are at the night when background light  levels are lowest. Another noise source comes from the detector characterised by one of the key parameters of the single photon counting module (SPCM), dark count rate (DCR). The system DCR is the sum of an intrinsic DCR due to the material properties of the detector, the biasing  conditions or the susceptibility to external noise while the $DCR_{b}$ is the registered event by the external noise photon that cannot be completely shielded~\cite{2009_Hadfield}. The intrinsic DCR increases exponentially when the bias current approaches the critical current but is below 1 Hz in the low bias region. The $DCR_{b}$ is around 10 Hz -1 kHz owing to the wide detection wavelength and high sensitivity of Superconducting nanowire single-photon detector (SNSPD)s~\cite{2015_Shibata}. Because the polarization of dark counting is independent on the incident signal, it contributes noise in the receiver~\cite{2022_Compagnino}. These external parts of noise are named as 

\begin{equation}
e_{E} = F(e_{A}, e_{N}+e_{DCR})
\label{eq:3}
\end{equation}

Where the $e_{A}$ is the part resulted from atmosphere, the $e_{N}$ is the part resulted from background noise and the $e_{DCR}$ is the noise comes from DCR. So, the whole QBER is defined as  

\begin{equation}
QBER = F(e_{I}, e_{E})
\label{eq:4}
\end{equation}

In QKD security analysis, any QBER must be considered as resulting from the activity of an eavesdropper, which requires post-processing to perform privacy amplification of the key based on the QBER to ensure adequate security. Therefore, efforts to reduce the QBER can increase the key generation rate, and the optimization process of the system focuses on the $e_{E}$ since the $e_{I}$ is only determined by the performance of the off-the-shelf source, and the off-the-shelf discriminating optics in receiver (specifically referring PBS here)~\cite{2022_Yan}. Bandpass filters are usually used to suppress background noise, which can greatly reduce all other noise beyond the band of the system signal. Generally speaking, the narrower filter window and the steeper cut-off edge, the better signal to noise ratio (SNR) could be achieved. However, considering the wavelength shift caused by the Doppler effect and temperature, the selection of the bandpass filter needs to be balanced between signal transmission and noise rejection to maximise the overall SKR. Besides, the background noise can be mitigated by gating or time-stamping the detection events. The minimum time  interval for gating or time-stamping is set by the timing jitter of the detector, beacon and time-to-digital converter.

As optical wavelengths are several orders of magnitude shorter than radio frequency signals, the diffraction limited divergence angles of 10 cm diameter beams are on the order of microradians. This brings optical communication the advantage of higher transmission efficiency compared to microwave communication, but it also leads to the need for an accurate pointing and tracking system to match the fields of view of satellites and OGS~\cite{2019_Chang}. In addition, due to the relative motion of satellites with respect to ground station and the complex and changeable atmospheric environment, traditional clock synchronisation schemes for QKD in free space will lead to  large jitter. The lack of synchronised high precision clocks results in inconsistencies between the transmitted and the received sequence and poor discrimination of background counts as a wide time gate is needed. Also, due to the continuous change of the satellite attitude relative to the OGS, providing an identical and stable polarisation reference is also essential to successful communication. Based on these considerations we have developed an elegant solution using a classical bright pulsed laser as beacon~\cite{2021_Zhang}, and here is the first time it has be used for QKD demonstration. The pulses are designed to be highly synchronised (less than 15 ps) with the quantum signal to indicate the exact time of quantum signal arriving. Each two adjacent beacon pulses represent a bit and the HDBC is modulated onto the pulses train to provide a unique sequence index for each arriving pulse. Finally, the beacon source adopts linearly polarized light with a sufficiently high extinction ratio and is consistent with the polarization state of the H channel, so as to provide a reference for the OGS. Polarisation matching through satellite attitude control can solve the above problems, but it requires accurate, stable and real-time closed-loop feedback control, which dramatically increases the complexity of the system and increases the fuel consumption of the satellite. Additionally it cannot compensate for the error caused by any atmospheric effects. The most widely used method is to polarisation compensate the received signal with a set of automatically rotating waveplates. Using the linear polarisation properties of the beacon light, an algorithm will control the automatic rotation of two quarter-wave plates and one half-wave plate until the intensity of the received light reaches its maximum\footnote{It is worth noting that because the beacon light and the signal light use different wavelengths, any birefringence caused by atmospheric effects is not independent of the wavelength, so it may lead to greater compensation errors in the signal at the receiving end.}.

\section{System and Experimental Result}

\begin{figure}[tbh]
\centering
\includegraphics[width=1.0\linewidth]{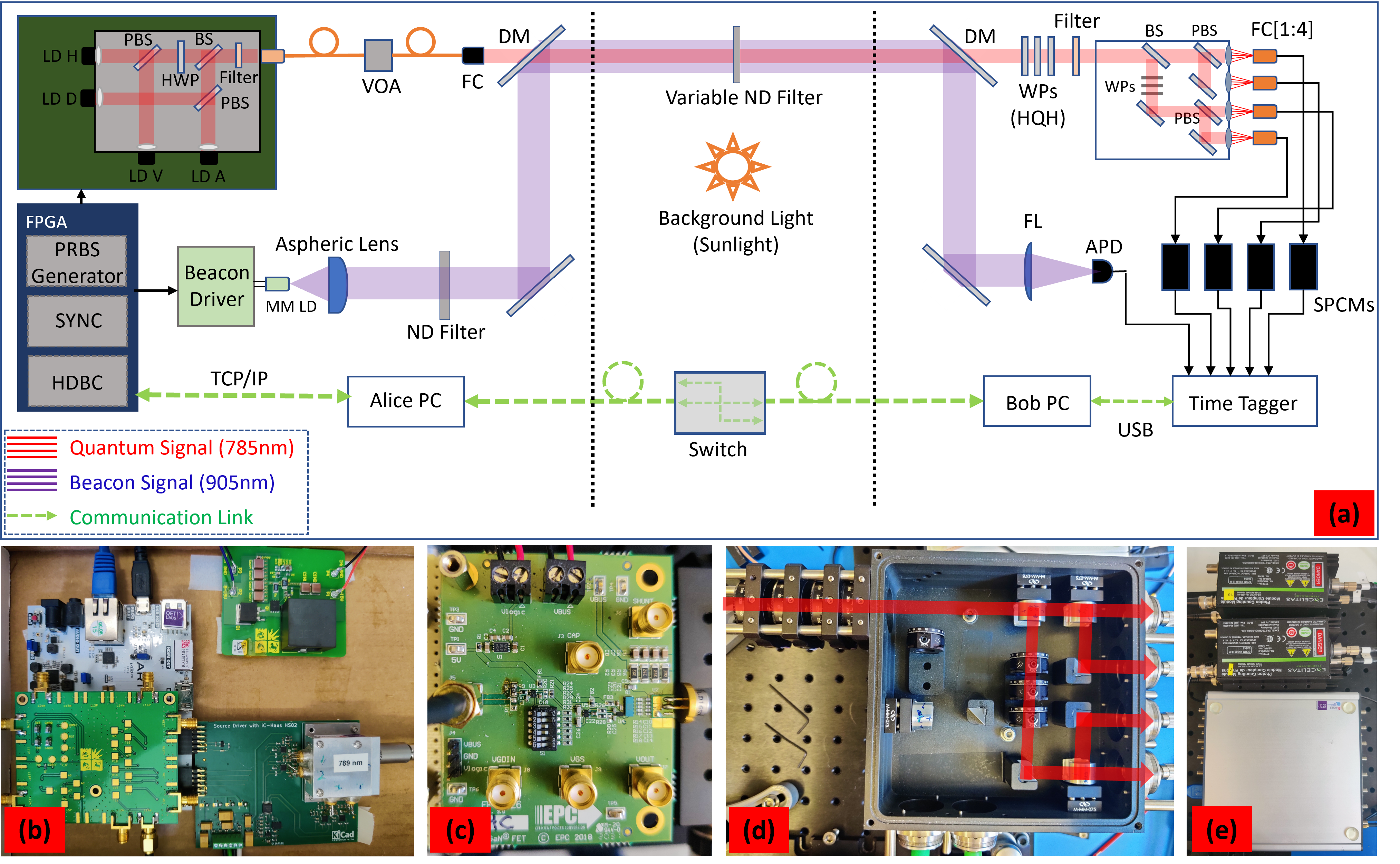}
\caption{End to End Satellite Free Space QKD Testbench. (a) The test platform is divided into three parts, with red parallel lines indicating the quantum signal at a wavelength of 785 nm, purple indicating the beacon optical channel at a wavelength of 905 nm, and green dashed lines indicating the distribution of the electrical signal. The middle is the communication channel. The right side is the receiver deployed on the ground. (b) is the field programmable gate arrays (FPGA) board and the adapted quantum source which is coupled by single mode fiber. (c) is the beacon driver which is triggered by FPGA, the laser diode is on the right edge of the board. The modulated pulses have a width of less than 4.1ns and a repetition of 100 kHz to achieve the goal of high SNR and low average power. (d) is the receiving box in Bob side, the top left 3 waveplates are used for polarisation compensation while the 4th is the bandpass filter to compress the noise. The optical path is marked using red line in the figure. (e) is the SPCMs (Excelitas SPCM-NIR) used for single photon detection, the bottom is the time tagger developed by Bristol which has 16 parallel channels with 1.6 ps resolution~\cite{2019_Tancock}.}
\label{fig1}
\end{figure}

This paper demonstrates a free-space QKD validation platform consisting of a transmitter (Alice) based on the BB84 protocol, a transmission channel with controlled background noise and loss, a receiver (Bob), and a classical communication link for completing post-processing. The platform is used to validate the functional operation of the QKD system in the laboratory, the performance at different losses and noise, and the post-processing algorithms. The transmitter used in the platform is the prototype for the ROKs mission (led by Craft Prospect Ltd) and a UK-Canada mission (piggy backed on the QEYSSAT mission). An upgraded version is being adapted for the UK EPSRC Quantum Communication Hub mission~\cite{2020_Mazzarella}, which is scheduled to launch into space for in-orbit demonstration in 2024. The receivers will be deployed on optical ground stations behind suitable telescopes depending on the mission. Fig.~\ref{fig1} shows the testbench diagram and specific design for the source and receiver. In (a) the left side is the transmitter, which consists of PC, FPGA, a quantum source, a beacon source and the corresponding optical path. The ZYNQ7000 FPGA provides a pseudorandom binary sequence (PRBS) generator for triggering the quantum source emitting the different polarisation states in a pseudo random sequence. The QKD source uses 4 individual laser diodes to emit pulses with different polarisation. A 2 nm bandpass filter is used to minimise the spectrum differences between different diodes to avoid potential security loopholes. The driver current is configured separately for each laser diode to make sure the intensity and pulse width is identical with a reasonable difference, as shown in Table.~\ref{tab1}. An HDBC encoder is implemented to trigger the beacon source generating a 100 kHz bright laser signal, which is  to achieve dual-channel synchronisation with jitter of less than 12 ps. A variable optical attenuator is connected to the quantum source via single mode fiber to set the $\mu$ which then is collimated into free space. The beacon source is designed based on a multimode laser diode with ultra high peak power up to 100 W. An aspheric lens is used for preliminary collimation and the ND filter is used for simulating the internal loss by aperture clipping due to the multimode and anisotropic beam. Finally a 820 nm cutoff dichroic mirror is used for combining the two beams (beacon and quantum) into a co-propagating beam. In the transmission channel, a variable ND filter is placed on the propagation path of the beacon to simulate the loss variations during the satellite's orbit. The background light in the system comes from the illumination of the sunlight and the intensity is controlled by a curtain. It should be noted that this is the prototype system demonstration before the development of flight model for final launch. We use it here mainly to prove operation at in high loss scenarios and subsequent secure key generation. Our ongoing device level work is proving that SWAP for the cubesatellite and other space environment requirements can be met~\cite{2023_Jaya}. The QKD payload shown in Fig.~\ref{fig1} (b) has three layers, including FPGA core board, source driving board and power board. Each layer has a area about 10cm x 10cm equivalently and the next integrated version has proofed that the whole payload can be fit into a 0.5 U cage.

\begin{table}[ht]
\centering

\caption{\label{tab2} The configurations for the End-to-End test}

\begin{tabular}{@{}lll}
\hline
Description & Parameter & Value \\
\hline
Quantum Source Repetition &           $R_Q$ &         25 MHz \\

Quantum Source Wavelength &     $\lambda_Q$ & 785 nm\\

Beacon Source Repetition  &           $R_B$ &         100 kHz\\

Beacon Source Wavelength  &           $\lambda_B$ &          905 nm\\

Intrinsic QBER            &           $QBER_i$ &       1.5\%\\

Dark Count (4 SPCMs)      &           $DCR$    &       2.3 kbps\\

System Loss (785 nm)       &      $L_{sys}$ &         7.4 dB\\

Detector efficiency (785 nm)       &      $L_{d}$ &         60\% (2.2 dB loss)\\

Mean Photon Number        &      $\mu$      &         0.1\\

Pulse Width (Max and Min)        &      FWHM      &         0.91 ns, 0.82 ns\\

\hline

\end{tabular}

\end{table}

In the receiving block located on the right, another 820 nm dichroic mirror is used to separate the quantum and beacon beams. The quantum signal is split into four channels by the receiver box and coupled to the corresponding SPCMs via a multimode fibre. A three waveplate array at the box entrance is used to correct for polarisation errors between the satellite and the OGS, and a bandpass filter with a transmission efficiency of 50\% suppresses all spectral light except the 785 nm with a 10 nm window to improve the SNR. Inside the receiving box, a beamsplitter (BS) splits the quantum signal into the two measurement bases, one arm of which passes through another waveplate used to rotate to the diagonal basis so that an H-V polarising BS can discriminate D-A. For the beacon detection, a detector circuit developed based on the Hamamatsu 
avalanche photodiode (APD) S13282 is deployed with a sensitivity of 4 MV/W and a noise level of 50 $fW/\sqrt{Hz}$ so that the faint pulse after up to 80 dB loss still can be detected. The two computers used to control Alice and Bob respectively are connected via TCP/IP protocol via a network switch for the necessary classical communication and post-processing to generate security keys. Table.~\ref{tab2} shows the parameters set in the system, which has some difference with the actual mission plan, but the result could be convert to the mission configuration as the estimation of proof-principle.

\subsection{Channel Loss Capability}\label{sec_loss}

\begin{figure}[htb]
\centering
\includegraphics[scale=0.4]{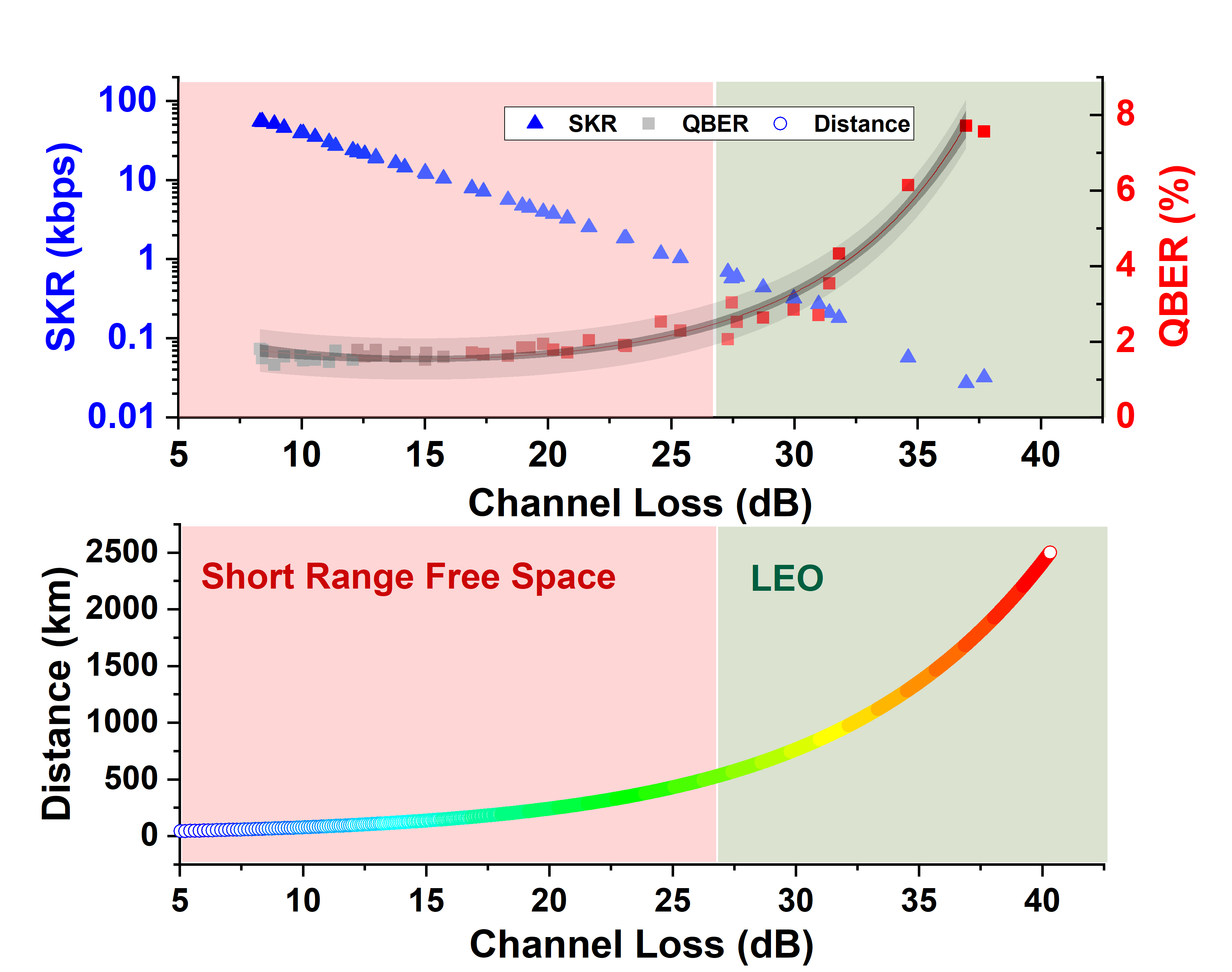}
\caption{Experimental result of SKR/QBER Vs channel loss. The system loss is 7.4 dB which includes all the components both in transmitter \protect\footnotemark[5] (1 dB) and receiver (6.4 dB). The additional channel loss is configured from 6.3 dB to 37.7 dB using a variable optical ND filter. Adding the system loss and the channel loss takes us to a maximum of 47.3 dB, which is beyond the maximum loss estimated in Section.~\ref{sec_2}. The figure shows the SKR change with the channel loss increase with the parameters configured as table.~\ref{tab2}. The initial trend before 30 dB is linear and then decreasing faster as the QBER increases rapidly~\cite{2022_Sidhu}. The background noise is 2.7 kHz in the experiment.}

\label{fig2}
\end{figure}

Considering a low-orbit satellite with an orbital altitude of 500 km, the geometric loss of the satellite to OGS in a clear atmosphere corresponds to 29.8 dB - 45.8 dB in the range of elevation angle larger than 10 degrees~\cite{2022_Sidhu}. Variations in channel loss affect the QBER due to fluctuations in the SNR on the one hand, and the  raw key due to changes in the number of arriving photons on the other, both of which affect the secure key rate and size. Figure.~\ref{fig2} shows the system performance under different channel loss which is quantified  by the SKR and QBER. In the experiment, a variable ND filter is mounted in the beam propagation path to simulate the channel loss, the beacon and quantum signals are combined via a dichroic mirror and simultaneously pass from the transmitter to the receiver through the fading channel.

\footnotetext[5]{In the real scenario, the mean photon number is set at the output aperture of the satellite telescope, so there is no extra loss introduced.}

In the top of Fig.~\ref{fig2}, SKR initially decreases linearly (on log-log scale) with the loss, but the rate becomes steeper after the loss reaches 30 dB. Likewise, the rate of increase in QBER also becomes steeper above losses of order 30 dB. By observing the SNR in different situations, it can be found that at the beginning due to the low loss, the SNR is high enough that the reduction of SKR only comes from the loss of the channel. When the loss reaches a certain value, the SNR is lower than a certain threshold, and the QBER increases sharply at this time, so that although the raw key still decreases linearly, the final SKR begins to decrease dramatically. Estimating system performance by SNR will be discussed separately in the final subsection. Here we distill the secure key by privacy amplification using the GLLP~\cite{2004_Gottesman} security proof and perform error correction using cascade error correction. Using low density polar code (LDPC) to do error correction will give a better yield of the secure key~\cite{2009_Elkouss}.

\subsection{Background Noise Effect}

\begin{figure}[htb]
\centering
\includegraphics[scale=0.4]{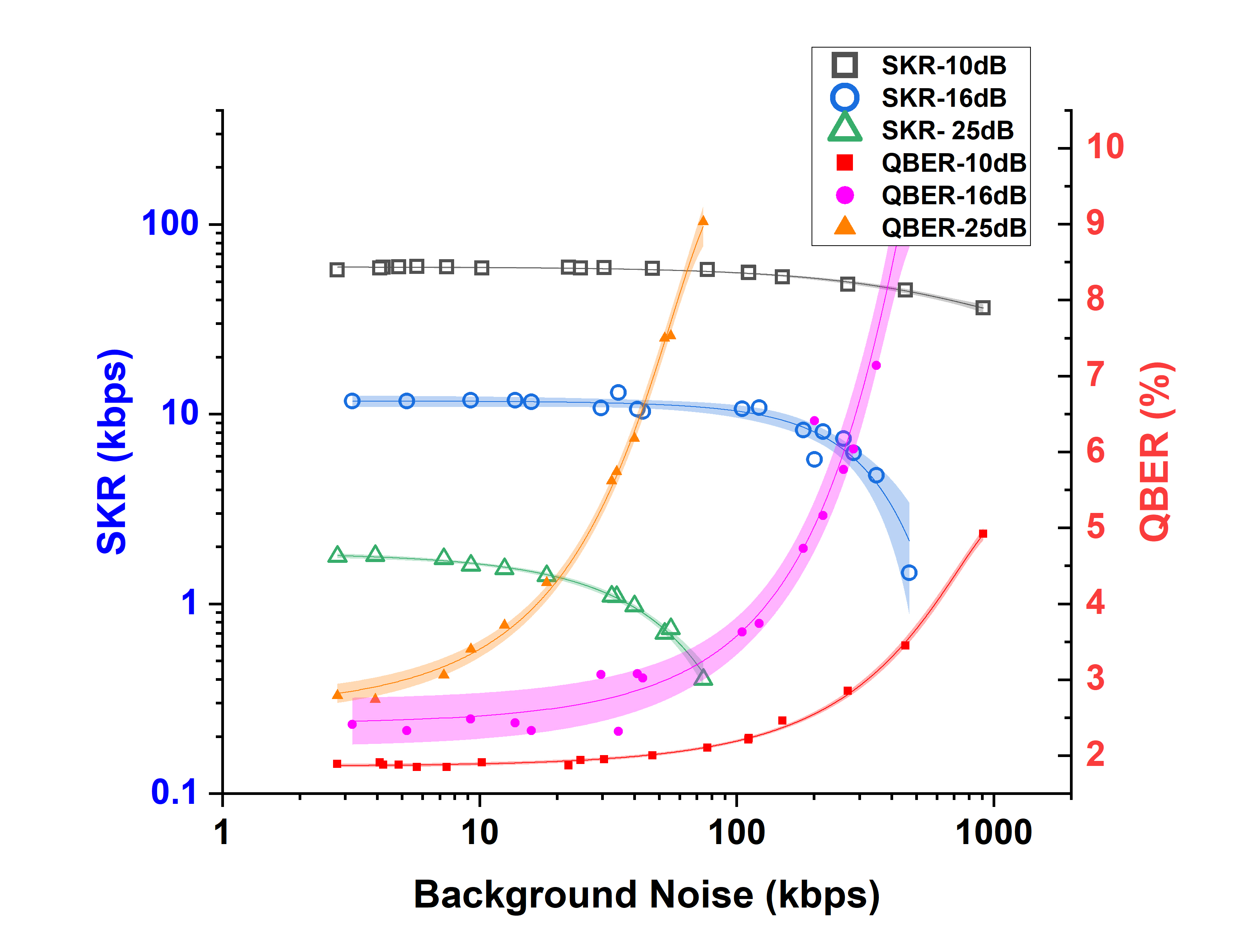}
\caption{Experimental result of SKR / QBER Vs background noise. The background brightness is characterized using the noise count rate of the SPCMs when sending no quantum light. The hollow legends represent SKR under different background light noise, and the solid legends represent QBER (shaded ares are indicative of the scale of error in parameters). The system performance using three different values of additional channel loss (10 dB, 16 dB, and 25 dB) is experimentally studied. System parameters are configured in table.~\ref{tab1}}
\label{fig3}
\end{figure}

The QBER and SKR under different background noise are investigated to see how the environmental light intensity will affect the performance. The background noise count rate is measured using the receiver unit while sending no quantum light. The background noise intensity is controlled by adjusting the light leakage into the (covered) experimental environment from external room/day light. In Fig.~\ref{fig3}, under the three different additional channel loss, SKR and QBER remain essentially unchanged initially. However, as the noise level further increases, QBER rises sharply after a certain point, accompanied by a sharp drop in SKR. Increased channel loss means that the noise level remains unchanged but the signal intensity is attenuated while here the noise level increases and the signal intensity remains unchanged. It is worth noting that the inflection point of SKR with different losses occurs at different background noise levels, 30 kbps, 200 kbps, 1000 kbps respectively. But this result is consistent with the trend of QBER and SKR when the loss increases in the Fig.~\ref{fig1} if considering from the point of view of SNR discussed in the section~\ref{sec_esnr}.

\subsection{Effect of Gate Width}

In the BB84 protocol, the quantum signals are pulse modulated, so photons are concentrated into  narrow periodic time windows characterised by full width half maximum (FWHM) below 1ns. To improve the SNR, a gating window is used to filter the signal received by the SPCMs, the click events inside the window will be kept and those outside will be removed. The selection and optimization of the gating window width to optimise SKR is discussed in the following. Taking a Gaussian signal pulse as an example, a wider gating window will collect more signals to form the raw key, but contain more noise and deteriorate the QBER. The crux of the matter is that the final secure key production rate depends on both parameters, which implies that there should be a certain window width with which the SKR is maximised.

\begin{figure}[htbp]
\centering
\includegraphics[scale=0.3]{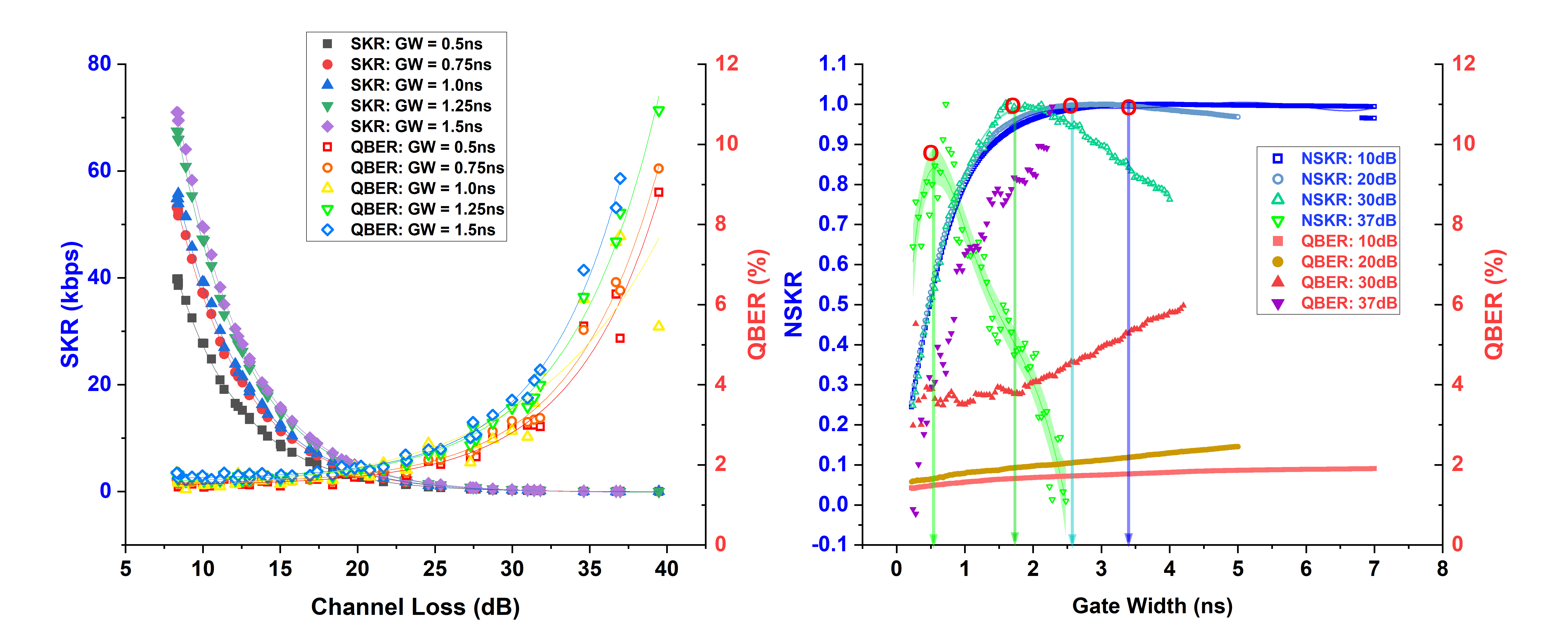}
\caption{Experimental result of gate width effect on SKR. The left figure shows the SKR and QBER change with channel loss using different gate widths to do postprocessing. The solid legends correspond to the SKR (left axis) while the hollow legends represents the QBER (right axis). The right figure shows the performance change under different gate widths. The hollow legends presents the normalised SKR (NSKR) while the solid legends present the QBER. The parameters used are configured in table.~\ref{tab2} and the extra background noise is 400 kHz except the DCR.}
\label{fig4}
\end{figure}

The speculation was proved by the corresponding experiment shown in Fig.~\ref{fig4}. The left figure shows the gating width effect on the SKR and QBER when the loss is changing. It's clear that the QBER with narrower $G_W$ is lower than that with wider $G_W$ under different channel loss. The secure key generated shows the opposite trend in that a wider $G_W$ get a higher SKR when the loss is less than 25 dB. This is because when the loss is not very high, a wider gate width provides a higher raw key rate without increasing the QBER too much, so the SKR manifests an increase correspondingly. However, things change when the loss is above a threshold which is explained by the right figure. The right figure shows the NSKR and QBER with different gating widths, the four data sets assume different channel loss, 10 dB, 20 dB, 30 dB, 37 dB. As SKR varies considerably at different losses, NSKR was adopted to characterise the rate of secure key generation in order to better represent the effect of gating width on the peak. The QBER with wider gating width has a sharper increase trend as the SNR is lower when the loss is higher. The NSKR shows four clear peaks when applying different gate width from 0.1 ns to 7 ns but the 37 dB peak happens at 0.55 ns while the 10 dB peak happens at 3.3 ns, this is because the introduction of noise in the presence of high losses is more likely to lead to a sharp drop in SNR.

\subsection{QBER and ESNR}
\label{sec_esnr}
The QBER is mainly affected by background noise, channel loss and $\mu$, and these effects are not independent of each other. From an in-depth analysis of equations~\ref{eq:1}-\ref{eq:4}, we can derive the following formula according to the definition of $QBER$.

\begin{equation}
QBER_{N} = \frac{N}{S+N} = \frac{1}{ESNR+1}  
\label{eq:5}
\end{equation}

\begin{equation}
ESNR = \frac{\int_{-\frac{G_W}{2}}^{+\frac{G_W}{2}} P(t)dt}{Nd} 
\label{eq:6}
\end{equation}

Where $N$ is the noise count, $S$ is the signal count, $GW$ is the gate width used for gating the signal, $P(t)$ is the function of the quantum pulse ($\mu=\int^\infty_{-\infty} P(t)dt $), $d$ is the duty cycle ($G_W\times R_Q$). Here we define ESNR as the effective SNR. SNR represents the signal to noise ratio in an average statistical time but not in a gate window, which isn't very accurate in this case. 

\begin{figure}[htbp]
\centering
\includegraphics[scale=0.5]{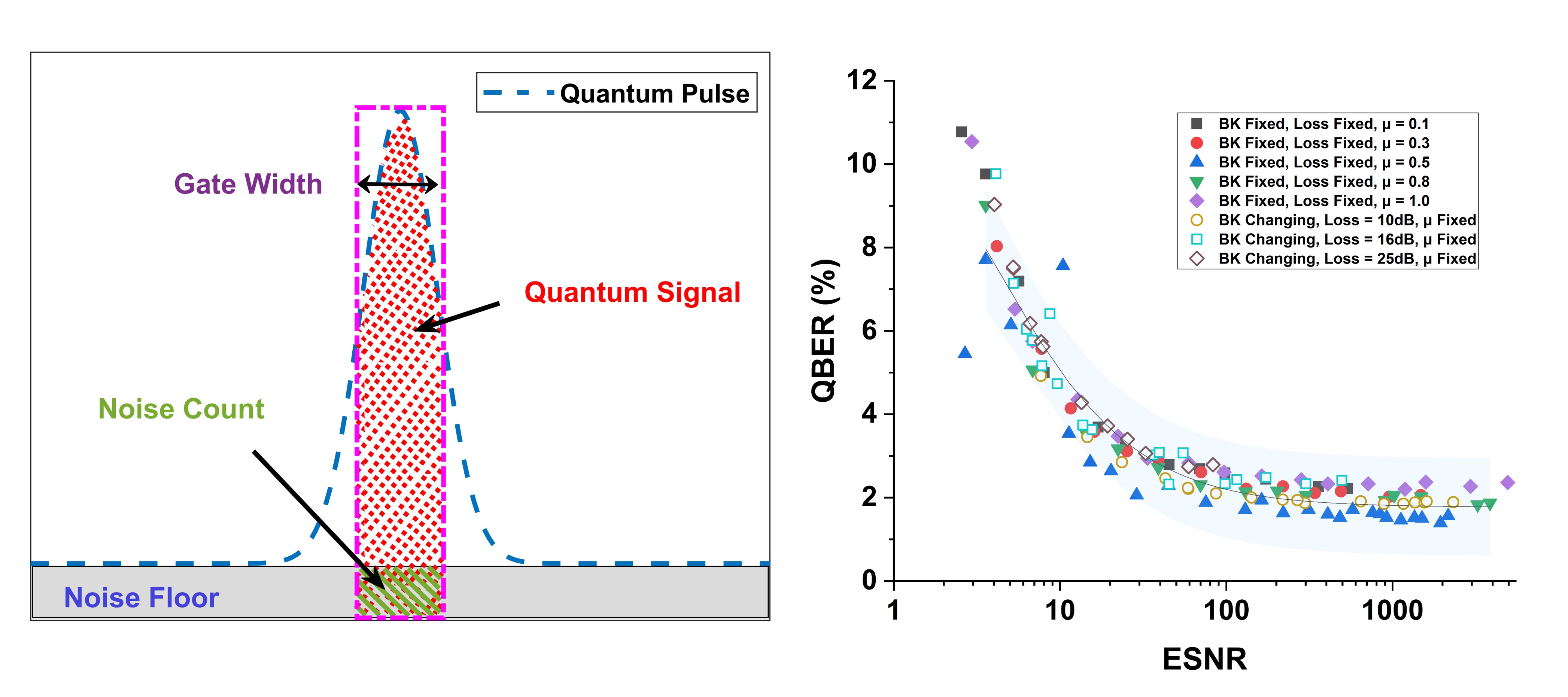}
\caption{Experimental result of QBER Vs ESNR. The ESNR combine the background noise (BK), $\mu$, channel loss and gate width together, which evaluates the performance of QKD more practically. The right figure shows the QBER from different cases (Different channel losses, background noise, and $\mu$) shows the same trends when increasing the ESNR.}
\label{fig5}
\end{figure}

From the Fig.~\ref{fig5}, the left figure shows the composition of the photon events registered by SPCMs, the respective contributions of signal and noise photons to them and the role of the gating window. When a narrow window is applied, less noise is introduced but more signal is lost and vice versa. The graph on the right depicts how the QBER grows for different system settings with variations in ESNR. In the first five curves, The background noise level and the channel losses are the same but the $\mu$ are set to 0.1, 0.3, 0.5, 0.8, 1.0 respectively\footnote[6]{It should be noted that the $\mu$ here is not set based on the theoretical optimised value for maximising the security key rate. Here we mainly want to show that the ESNR is more reliable for qualifying the performance so that the $\mu$ value is picked uniformly in the range of 0.1 - 1.0 from the engineering point of view.}. the 6 - 8 curves show the experimental result with different background noise, different channel losses (10 dB, 16 dB, 25 dB) and same $\mu = 0.1$. With the equation~\ref{eq:6}, the QBER keep the same falling trend with the ESNR increasing.  This gives an interesting point that when analysing the effect on QBER from different losses, $\mu$, background noise, gating widths, each parameter produces a different trend. However, when we transform all the above parameters into ESNR and look at it's relationship with QBER, all these points from different data sets (with different parameters) obey the same variation. This means that using ESNR to estimate and evaluate QBER and SKR can lead to more valid conclusions.

\section{Improvement and Optimisation}

\textbf{Loss budget}. In the quantum channel of the end-to-end testbench, the finite optical path length introduces no geometric loss to the system, the 7.4 dB attenuation is caused by the fiber coupling efficiency and the bandpass filter's transmission, called $\eta_I$. The narrowband filter has a transmission of 50\% for 785 nm quantum signal, contributing 3 dB loss. The remainder comes mainly from free space-to-multimode fibre coupling at the output of the receiver box and a pair of dichroic mirrors with a transmission of approximately 80\%. The maximum channel loss added in the experiment is 37.3 dB, considering the 2.2 dB (60\%) detection efficiency of the SPCMs, the total loss that can be tolerated is 46.9 dB in this work. In the real scenario, the maximum geometric channel loss is about 27.7 dB with a 10 microrad divergence and 70 cm telescope aperture. With 7.9 dB absorption loss, the total loss budget with the current setup is $Loss_{Q} = 45.8 dB$ (SPOQC Mission). The improvement of bandpass filter transmission in the OGS from 50\% to 90\% will reduce the aggregated loss by 2.6 dB. For the beacon and synchronisation channel, the maximum loss budget is 79.4 dB for the ROKs cube satellite mission due to a larger divergence (100 microrad) as it is also used for the pointing and tracking, while it is only 50.0 dB for the hub mission (synchronisation only). In this test, the beacon not only passes through the same ND filter with the quantum beam, but also through an extra 42.2 dB ND filter (calibrated at the wavelength of transmission). The dichroic mirror pair and focusing lens in front of the APD contribute another 3 dB. So the total loss suffered in this experiment is 85.2 dB, which clearly suggests it can cope with operation in orbit for the ROKs mission. The SPOQC mission will utilise a single mode laser which can provide a 400 mW peak power but with a small divergence of 18.7 microrad so that the received intensity could provide enough SNR ($>10$) for timing and synchronisation.

\textbf{Mean photon number}. Note that here we have tested QKD using a $\mu = 0.1$. However, we intend to implement the decoy state protocol where the secure key can be generated with a higher effective photon number.  The optimal intensity parameters have been estimated in ~\cite{2022_Sidhu}. Here we use representative values of $\mu _1= 0.5$, $p_1= 0.72$, $\mu _2= 0.08$, $p _2= 0.18$, $\mu _3= 0$, $p _3= 0.1$, with the transmitter and receiver basis bias set to $P_X= 0.9$, $P_Z = 0.1$. In this optimal system an overall $\mu = 0.3744$ can be obtained and the reconciliation ratio could be up to 0.82 from the asymmetric basis selection, which will give another 7.9 dB raw key rate gain.

\textbf{Dark count}. The DCR of the detector used here is 500 Hz, and using SPCMs with DCR $<$ 100 Hz will result in a 7 dB increase in SNR. The reduced DCR will further decrease the QBER and increase the secure key rate.

\textbf{Equivalent finite key length in real scenario}. LEO's intermittent communication window allows a limited length of key to be generated per satellite pass, which exacerbates the generation of secure keys as a significant portion of received signal must be sacrificed for accurate parameter estimation while retaining enough raw key for there to be a finite secure key after postprocessing. Considering the $QBER_I$ = 0.5\%, the finite key length is about 6.5 Mbits with the elevation angle larger than 10 degrees.~\cite{2022_Sidhu}. The aimed $R_Q$ is 400 MHz which could get at least a 12 dB SKL improvement over this 25 MHz system (assuming comparable $QBER$). Improving the quantum efficiency of the detector is another way to improve SKL but wouldn't be too much.

\begin{table}[ht]
\centering

\caption{\label{tab3} The parameter improvement in mission target}

\begin{tabular}{@{}lllll}
\hline
Description & Parameter & This Work & Mission Target & Gain\\
\hline
Quantum Source Repetition & $R_Q$ & 25 MHz & 400 MHz $\uparrow$ & 12 dB\\

Intrinsic QBER & $QBER_i$ & 1.5\% & 0.5\% $\downarrow$ & $\uparrow$ \\

Dark Count (4 SPCMs) & $DCR$ & 2.3 kbps & 2.3 kbps & 0 dB\tablefootnote[7]{This parameter should have improvement in the final system by replacing better component, but we leave them as the potential loss tolerance margin.}\\

System Loss (785 nm) & $L_{sys}$ & 7.4 dB & 3.8 dB $\downarrow$ & 3.6 dB\\

Detector efficiency (785 nm) & $L_{d}$ & 60\% (2.2 dB loss) & 60\% (2.2 dB loss) & 0 dB\footnotemark[7]\\

Mean Photon Number & $\mu$ & 0.1 & 0.3744 $\uparrow$ & 5.7 dB \\

\hline

\end{tabular}

\end{table}

\begin{figure}[htbp]
\centering
\includegraphics[width=0.6\linewidth]{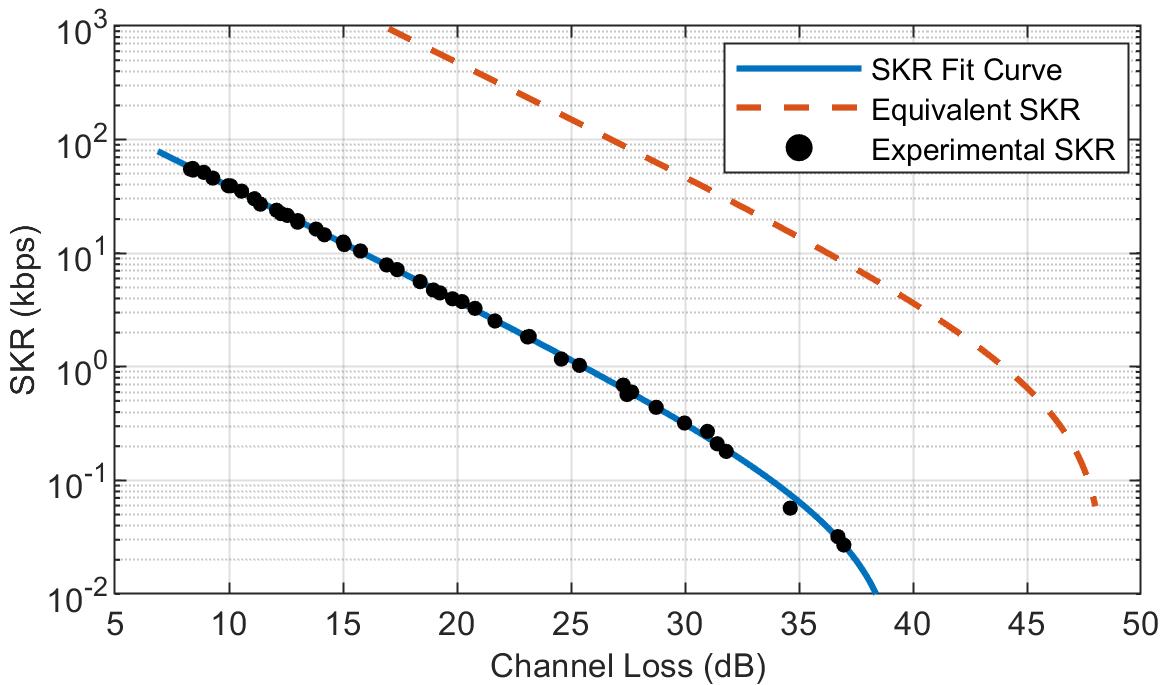}
\caption{Experimental result of SKR and the Equivalent SKR in the target mission. The black points shows the SKR evaluated in the experiment with a blue fit curve, and the orange dash line shows the estimated equivalent SKR when converting the experiment loss budget to the targeted mission budget. This involves shifting the curve horizontally by 9.3 dB taking into account reductions in system loss and increased $\mu$ and moving vertically by 12 dB to account for increases in repetition rate.}
\label{fig6}
\end{figure}

In the comparison between this work and the mission target, there is a gain of 21.3 dB that can be achieved  using the higher source repetition rate, higher $\mu$ (protected by decoy states) and lower system loss. Although we plan to use lower dark count detectors we also expect an increase of background light when outside the laboratory thus we expect broadly similar dark plus backround of around 2.3 kcps. Also the intrinsic QBER is expected to be slightly better but we will not consider it here as it's non-linear effect on the SKR. With these assumptions we can  project the result from this work to estimate the equivalent SKR in the real mission, as shown in Fig.~\ref{fig6}. Using the gains in $\mu$ and reduced losses we can shift the maximum loss in Fig.~\ref{fig2} by 9.3 dB (without changing the shape of the curve) from 37.7 dB to 47 dB  easily exceeding the maximum expected mission loss of 44.6 dB from table \ref{tab1}. The gain of 12 dB from increased repetition rate simple increases the key rate shifting the predicted key rate curve upwards. Although we only consider asymptotic key rates in this paper, the increase total key generated per satellite pass will increase our chances of also satisfying finite key requirements of security.  As a comparison between the fiber link and free space link at the same distance of 1000 km, the SKR in satellite QKD is expected to be 3.47 x $10^{-4}$ per pulse, while in the Twin-Field QKD the SKR is 9.53 x $10^{-12}$ due to the extremely high loss of 156.5 dB~\cite{2023_Liu}.

\section{Conclusion}

A satellite-based QKD payload and receiver testbench for proof-of-principle has been demonstrated and the impact of a range of system parameters on performance has been investigated in detail. This work demonstrates the feasibility of the design solution and provides valid references and recommendations for free-space QKD-based applications. In the next step, we will integrate the payload section to fit the mechanical profile of the satellite including integrating the electronics and optical path of the beacon light, upgrading the quantum source to achieve dual-wavelengths with a repetition of 400 MHz and implementing decoy state-BB84. At this point we expect to produce a working 0.3U flight model early in 2024\cite{2023_Jaya}. The internal losses of the system will also be optimised to improve the final SKR. The discussion offers valuable insight not only for the design and optimization of miniature low-cost satellite-based QKD systems but also any other short range free space QKD on the ground or in the air~\cite{2020_Liu}.

\section{Acknowledgments}

This research was funded by the EPSRC Quantum Communications Hub (EP/T001011/1), and the UK Space Agency (NSTP3-FT2-065 QSTP: Quantum Space Technology Payload, NSIP-N07 ROKS Discovery) and Innovate UK project 78161: ReFQ. The first author was supported by the University of Bristol - China Scholarship Council joint-funded Scholarship to participate in this research. The authors acknowledge discussions with Dr Daniel Oi and Dr Jasminder Sidhu on simulation of losses during a satellite pass and optimal parameters for decoy state QKD.

\clearpage

\appendix

\section{Channel Loss Estimation}

\begin{figure}[htbp]
\centering
\includegraphics[width=0.7\linewidth]{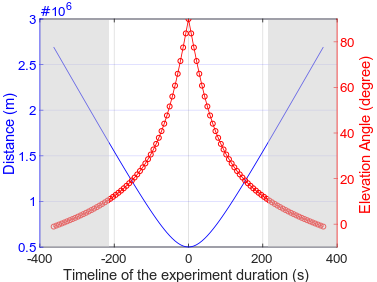}
\caption{The figure shows the distance and elevation angle change during the visibility window. The circle line shows the Elevation angle while the blue solid line shows the distance. The result is modeling based on the 3D Doppler model for the LEO~\cite{1998_Ali}. The result is calculated with the value in the table~\ref{tab3}. }
\label{fig7}
\end{figure}

Here, the channel loss modeling is given based on the diffraction limit as shown in Fig.~\ref{fig7}. According to the analysis of the visibility window and the link efficiency, the elevation angle is configured larger than 10 degrees. The corresponding minimal and maximal distance is 500 km and 1700 km. Based on the profile of the Gaussian beam, the diffraction limited half angle with the threshold of $1/e^2$ is given by

\begin{equation}
\theta _{0}=\frac{\lambda }{\pi \omega _{0}}M^{2}
\label{eq:7}
\end{equation}

Where the $\lambda$ is the wavelength and the $\omega _{0}$ is the radius of the beam wist. 

After propagation during the free space, the footprint in front of the telescope is

\begin{equation}
R\left ( d \right )=\tan \theta _{0}\times d
\label{eq:8}
\end{equation}

Where the $d$ is the distance between the satellite and the OGS, and the optical intensity of the beam at this distance is still Gaussian distribution given by

\begin{equation}
I=I_{0}e^{\frac{-2r^{2}}{R^{2}}}
\label{eq:9}
\end{equation}

Assuming that the centre of the spot coincides with the centre of the telescope aperture, the channel efficiency can be given by integrating the intensity over the aperture area

\begin{equation}
T=\frac{\int_{0}^{D_{T}/2}e^{\frac{-2r^{2}}{R^{2}}}}{\int_{0}^{\infty }e^{\frac{-2r^{2}}{R^{2}}}}=1-e^{\frac{-D_{T}^{2}}{2R^{2}}}\approx \frac{D_{T}^{2}}{2R^{2}}
\label{eq:10}
\end{equation}

\end{document}